\documentclass{aa}
\usepackage{graphics}

\usepackage{epsfig}
\usepackage{latexsym}
\usepackage{natbib}

\def\E$\gamma${E_\$\gamma$}

\def\deg       {$^{\circ}$}

\def \gray     {$\gamma$-ray}
\def \grays    {$\gamma$-rays}

\begin{document}

\title{An Unidentified Variable Gamma-Ray Source near the Galactic Plane
 Detected by COMPTEL
}

\author{ S.~Zhang\inst{1,2},
	W.~Collmar\inst{1},
         V.~Sch\"onfelder\inst{1}
}

\offprints{S.~Zhang} 
\institute{Max-Planck-Institut f\"ur extraterrestrische Physik,
               P.O. Box 1603, D-85740 Garching, Germany 
	     \and
	      High Energy Astrophysics Lab, 
              Institute of High Energy Physics,
	     P.O.Box 918-3, Beijing 100039, China
          }

\mail{szhang@mail.ihep.ac.cn}

\date{Received May 13, 2002 / Accepted October 1, 2002}

\titlerunning{COMPTEL Observations ...}
\authorrunning{S.~Zhang et al.}

\abstract{
We report the detection of an unidentified $\gamma$-ray source 
near the Galactic plane by the COMPTEL experiment aboard the 
Compton Gamma-Ray Observatory. 
The source is detected at a significance level of $\sim$ 7.2$\sigma$
in the energy range 1-3~MeV and at $\sim$~4.6$\sigma$ in the lower
0.75-1~MeV band in the time period March to July 1995.
At energies above 3~MeV are only marginal hints 
or upper limits obtained.  The MeV spectrum has a soft shape.
Strong flux variability is found within one year at energies below 3 MeV.
Possible counterparts of galactic and extragalactic nature are discussed. 


}
\maketitle

\section{Introduction}

The Compton Gamma-Ray Observatory (CGRO), carrying 4 \gray\ experiments
(BATSE, OSSE, EGRET, COMPTEL), has observed the universe
in $\gamma$-ray energies  with unprecedented
sensitivity between its launch in April 1991 and its reentry into the 
earth atmosphere in June 2000.
These observations provided many new and exciting results, particularly
the detection of many $\gamma$-ray point sources.
Surprisingly, a big fraction of them is still unidentified.
Among the 271 EGRET $\gamma$-ray sources listed in the 3rd EGRET catalogue 
for energies above 100~MeV, 171 are unidentified  (Hartman et al., 1999).
Some of them, located at low latitudes ($|$b$|$ $<$ 10\deg), show a strong positional
correlation with SNRs and OB associations (e.g. Romero et al., 1999),
while some others ($|$b$|$ $<$ 10\deg) do not coincide with any known
potential $\gamma$-ray emitters (Torres et al., 2001a). This absence of positional coincidences 
might indicate the presence of a new class of $\gamma$-ray sources 
(Torres et al., 2001a). 

COMPTEL was sensitive to soft $\gamma$-rays 
(0.75-30 MeV) and finally opened the MeV band - which basically was
unknown prior to CGRO - 
as a new astronomical window. Apart from $\gamma$-ray bursts, 
unidentified sources and AGN are the majority of the COMPTEL detections.
While the sum of all others (pulsars, stellar black-hole candidates,
supernova remnants and  $\gamma$-ray line sources) are about 12,
the first COMPTEL catalogue lists 10 AGN and 9 unidentified $\gamma$-ray sources
(Sch\"onfelder et al., 2000).  In this paper we report 
the detection of a further  unidentified  COMPTEL source, which shows 
a variable MeV flux and is located in the galactic plane.

\section{Instrument and Data Analysis}
The imaging Compton Telescope COMPTEL is sensitive to $\gamma$-rays in 
the energy range 0.75-30~MeV with energy-dependent energy and angular resolution
of 5$\%$ - 8$\%$ (FWHM) and 1.7\deg - 4.4\deg (FWHM), respectively.
Imaging in its large ($\sim$1 steradian) field of view is possible 
with a location accuracy of the order of 1$^{\circ}$-3$^{\circ}$ depending 
on source flux. The effective area of COMPTEL drops with increasing offset 
to the pointing direction. At an offset angle of $\sim$30\deg\ it is decreased to 
$\sim$50\% (energy dependent). 
For details on the COMPTEL experiment see Sch\"onfelder et al. (1993).

Skymaps and source parameters, like detection significances, fluxes and flux errors,
can be obtained via the maximum likelihood method, which is implemented 
in the standard COMPTEL data analysis package.
The detection significance is derived from the 
quantity -2ln$\lambda$, where $\lambda$ is the ratio of the likelihood L$_{0}$
(background) and the likelihood L$_{1}$ (source + background). 
The quantity -2ln$\lambda$ has a $\chi_{3}^{2}$ (3 degrees of freedom) distribution
for a source search and a $\chi_{1}^{2}$ distribution for a flux estimate of a 
known source (de Boer et al., 1992). 
An estimate for the instrumental background of COMPTEL is derived by using
the standard filter technique in the COMPTEL data space (Bloemen et al., 1994). 
In the presented analysis we applied instrumental point spread functions 
 assuming an E$^{-2}$ power law shape for the source spectrum.
The nearby known COMPTEL source, the $\gamma$-ray pulsar PSR 1509-58 (l/b = 320.3$^{\circ}$,-1.2$^{\circ}$), 
has been included in the analyses, i.e. its fluxes were estimated 
simultaneously with the ones of the new source by a fitting process, 
and have been subtracted off in the maps of Fig.~1.
The same procedure was applied for the celestial background
components, the galactic and extragalactic diffuse gamma-ray radiation.
\begin{table*}[thb]
\caption{COMPTEL observations of the sky region of the unidentified 
\gray\ source during its high state in 1995 (CGRO Phase 4).
The CGRO VPs, their time periods in calendar date and TJD, prime observational
targets of CGRO, pointing offset angles and effective exposures are given.}
\begin{flushleft}
\begin{tabular}{cccccc}
\hline 
\multicolumn{1}{c}{VP}&\multicolumn{1}{c}{Date}&\multicolumn{1}{c}{TJD }&\multicolumn{1}{c}{Object}&\multicolumn{1}{c}{Offset angle}&\multicolumn{1}{c}{Effective exposure }\\ 
\multicolumn{1}{c}{}&\multicolumn{1}{c}{(dd/mm/yy)}&\multicolumn{1}{c}{}&\multicolumn{1}{c}{}&\multicolumn{1}{c}{}&\multicolumn{1}{c}{days}\\ \hline 
414.0 & 21/03/95-29/03/95 &9797-9805&  Vela Pulsar & 33\deg& 0.55\\  
414.3 & 29/03/95-04/04/95 &9805-9811& GRO J1655-40 & 35\deg& 0.58\\ 
415.0 & 11/04/95-25/04/95 &9818-9832& LMC & 42\deg& 0.67\\ 
421.0 & 06/06/95-13/06/95 &9874-9881& Gal. Center & 43\deg& 0.28\\ 
422.0 & 13/06/95-20/06/95 &9881-9888&  Gal. Center & 43\deg& 0.27\\ 
423.5 & 30/06/95-10/07/95 &9898-9908& PKS 1622-297 & 36\deg& 0.70\\ 
424.0 & 10/07/95-25/07/95 &9908-9923& Cen A & 20\deg& 2.39\\ 
\hline
\end{tabular}\end{flushleft}
\label{tab:obs}
\end{table*}

\section{Observations}
CGRO observations were organized in so-called CGRO `Mission Phases' and `Viewing Periods'
(VPs). A `Phase' covers typically 1 year and contains many VPs,  
which typically last for 1 to 2 weeks (see e.g. Table~\ref{tab:obs}).
We have analyzed 48 such VPs, which are selected to be between the
beginning of the CGRO mission in 1991 and the second reboost of CGRO in 
March 1997 (within CGRO Phase 6), and which have pointing offsets between instrument and 
$\gamma$-ray source (l/b=311.5$^\circ$, -2.5$^\circ$) of less than
50 degrees. These observations are located in the first five CGRO Phases
and sum up to an effective exposure (100\% COMPTEL
directly on target) of 35.26 days on the source position.
\begin{table*}[tbh]
\caption{Fluxes of the unknown \gray\ source, the time periods and the effective
 exposures for different observational combinations are listed.
No source observation in CGRO Phase 6 before the second 
reboost of CGRO is carried out. The error bars are 1$\sigma$ and the upper limits
are 2$\sigma$.}
\begin{flushleft}
\begin{tabular}{ccccccc}\hline
\multicolumn{1}{c}{Period}&\multicolumn{1}{c}{TJD}&\multicolumn{1}{c}{Effective exposure }&\multicolumn{4}{c}{Flux (10$^{-5}$ ph cm$^{-2}$ s$^{-1}$)}\\  
\multicolumn{1}{c}{ }&\multicolumn{1}{c}{} &\multicolumn{1}{c}{days}&\multicolumn{1}{c}{0.75-1 MeV}&\multicolumn{1}{c}{1-3  MeV}&\multicolumn{1}{c}{3-10 MeV}&\multicolumn{1}{c}{10-30 MeV}\\  \hline
Phase 1 &8392-8943      &9.29   &6.0$\pm$3.1   &$<$6.0   &1.7$\pm$1.6   &0.8$\pm$0.6 \\ 
Phase 2 &8943-9216      &5.87   &4.2$\pm$3.5   &$<$10.1   &$<$3.9   &1.8$\pm$0.9 \\  
Phase 3 &9216-9629      &7.45   & $<$12.5   &6.7$\pm$3.8   &2.2$\pm$1.6   &$<$1.6 \\ 
Phase 4 &9629-9993      &8.42   &13.7$\pm$3.8   &21.2$\pm$3.5   &4.6$\pm$1.6   &$<$1.2 \\ 
Phase 5 &9993-10371     &4.23   &18.6$\pm$4.7   &$<$14.7   &$<$4.8   &$<$2.6 \\  
VPs 414-424 &9797-9923  &5.44   &19.9$\pm$4.6   &36.6$\pm$4.7   &5.4$\pm$2.3   &$<$1.8 \\ 
Phases 1-5  &8392-10371 &35.26  &8.4$\pm$1.7   &6.1$\pm$1.6   &1.6$\pm$0.8   &0.5$\pm$0.3 \\  \hline
\end{tabular}\end{flushleft}
\label{tab:flux}
\end{table*}
\begin{figure}[tb]
\centering
\psfig{figure=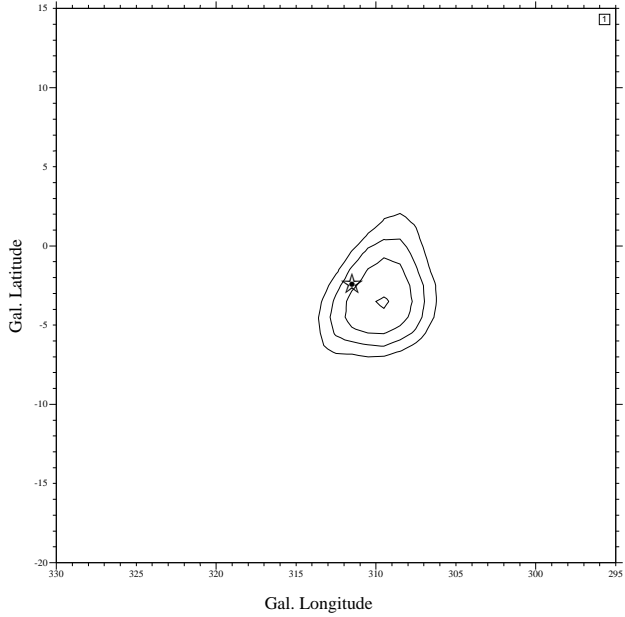,height=9cm,width=9.0cm,clip=}
\psfig{figure=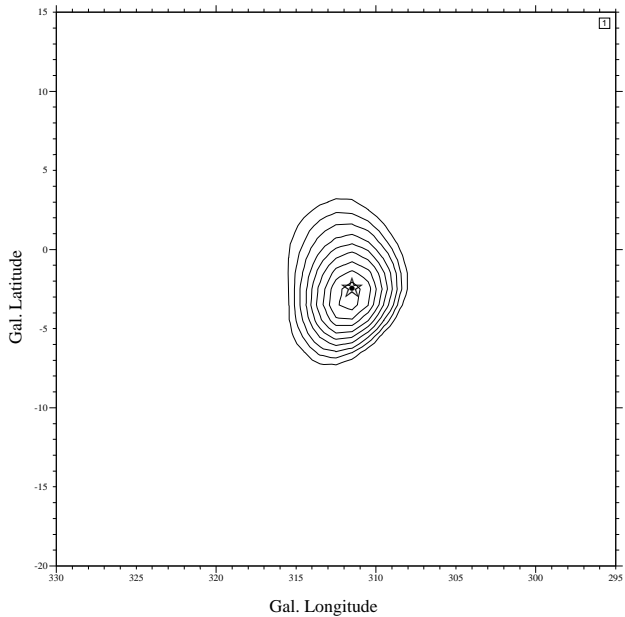,height=9cm,width=9.0cm,clip=}
\caption{Maps of the sky region of the unknown MeV source 
in the 0.75-1~MeV band ({\it{top}}) and the 1-3~MeV band ({\it{bottom}}) for VPs 414-424.
The star ($\star$) represents the most likely source position.
The known MeV source PSR 1509-58 as well as the diffuse \gray\ emission are 
subtracted off from the maps.
The contour lines start at a detection significance level of 3$\sigma$ with steps of
0.5$\sigma$.}
\label{fig:map}
\end{figure}
\begin{figure}[tb]
\centering
\psfig{figure=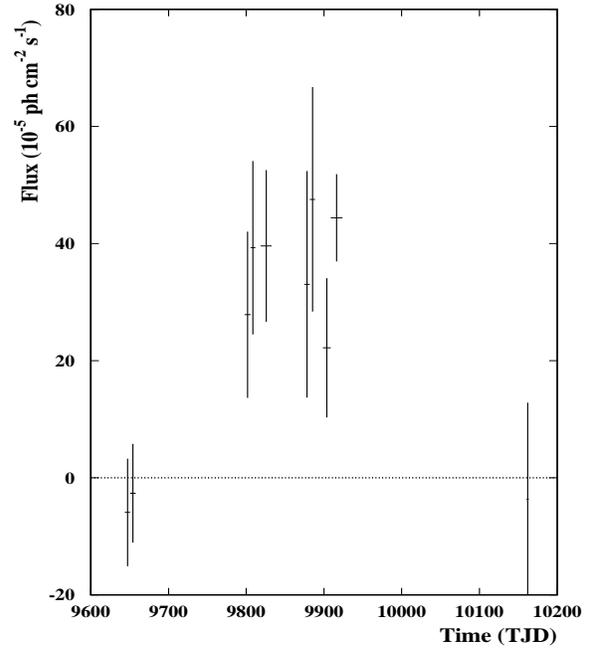,height=9cm,width=8.0cm,clip=}
\caption{1-3~MeV light curve of the unknown \gray\ source between 
October 1994 - March 1996. This period covers the outburst.
Each bin represents an individual CGRO VP. The error bars are 1$\sigma$.}
\label{fig:light-p4}
\end{figure}

\begin{figure}[tb]
\centering
\psfig{figure=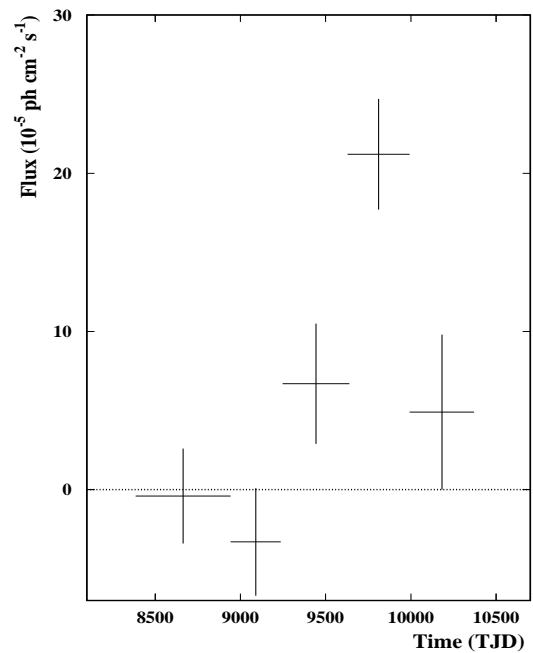,height=9cm,width=8.0cm,clip=}
\caption{Light curve of the unknown \gray\ source in the 1-3 MeV band. Each bin is averaged over an individual
CGRO Phase up to Phase 5. The error bars are 1$\sigma$.}
\label{fig:light-p15}
\end{figure}

\begin{figure}[tb]
\centering
\psfig{figure=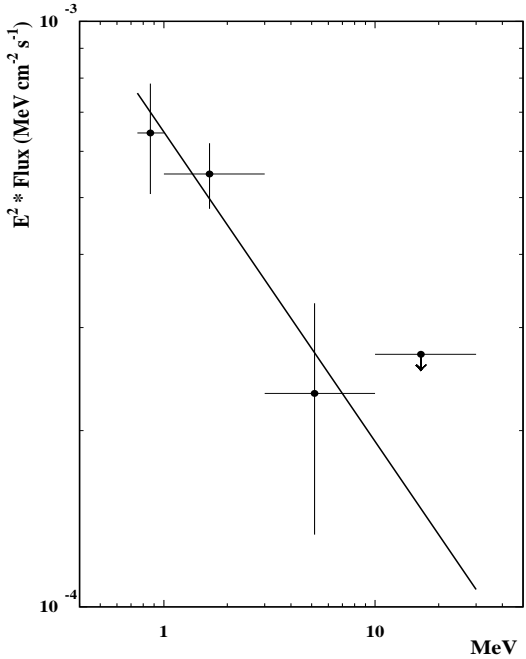,height=9cm,width=8.0cm,clip=}
\caption{The energy spectrum of the unknown \gray\ source for
the combination of 7 VPs (414-424) in Phase 4. The solid line shows
the best-fitting power law shape for the combination of VPs.
The error bars are 1$\sigma$ and the upper limits are 2$\sigma$.}
\label{fig:spectra}
\end{figure}

\begin{figure}[tb]
\centering
\psfig{figure=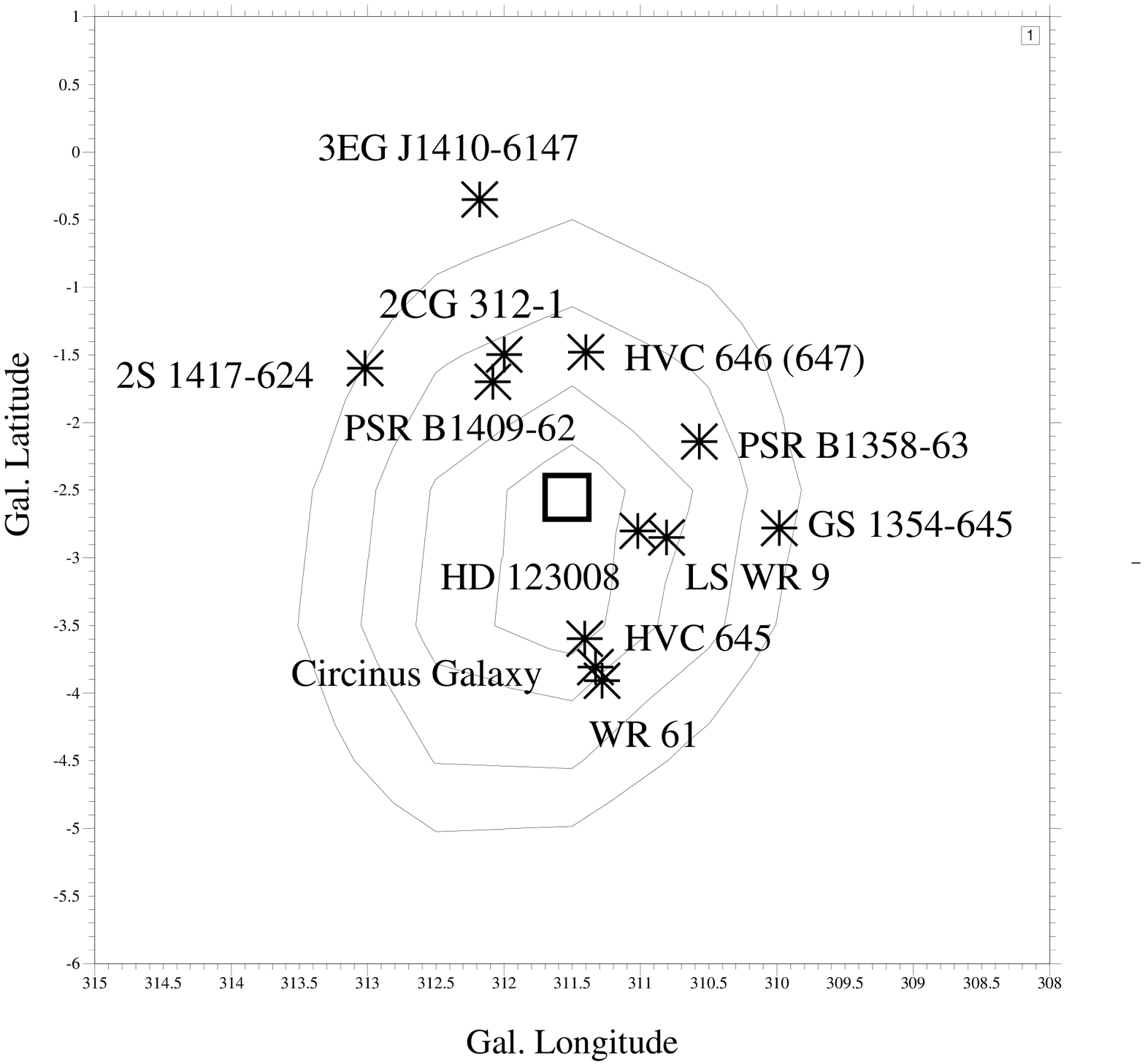,height=9cm,width=8.0cm,clip=}
\caption{Error location contours for the unknown \gray\ source for
the most significant detection: the 1-3 MeV band for the combination of 7 VPs
 (414-424). The most likely source position ($\Box$) and possible counterparts 
are overlaid. The error contours start at 1$\sigma$ with steps of 1$\sigma$. }
\label{fig:counterpart}
\end{figure}

\section{Results}
\subsection{Detections}
Evidence for the MeV source was first found by analyzing 
CGRO VP 424 (July 10-25, 1995) where the radio galaxy Centaurus A 
was the prime CGRO target. In the 1-3~MeV band a bright source 
with a detection significance of 5.2$\sigma$ is visible at a
position near l/b = 311.5$^\circ$, -2.5$^\circ$. In addition, hints for
this source were found in each of the 6 previous VPs covering this sky 
region. By combining these 7 VPs, which cover the time period March 21 
to July 25 in 1995 (Table~\ref{tab:obs}), we derive the
best detection significances of 7.2$\sigma$ in the 1-3 MeV band
and of 4.6$\sigma$ in the 0.75-1 MeV
band by assuming 3 degrees of freedom (Fig.~\ref{fig:map}).
The probability of detecting randomly an unknown source with a
-2ln$\lambda$ value of 60 (corresponding to 7.2$\sigma$) 
in the COMPTEL 1-3~MeV band is 5.4$\times$10$^{-10}$ by taking
into account the trials for searching in all individual VPs 
in four energy bands from the 
beginning of CGRO mission in 1991 to its second reboost in 1997.
Above 3~MeV the evidence for the source becomes marginal. 
The source fluxes in the 7 VPs are roughly the same
(Fig.~\ref{fig:light-p4}). The detection significance is largest in
VP 424 due to the best exposure (Table~\ref{tab:obs}). 
The derived source fluxes are given in Table~\ref{tab:flux}.
By combining the observations of the first five CGRO Phases, we derive 
time-averaged significances at energies below 3 MeV (3 dof) of
4.7$\sigma$ in the 0.75-1~MeV band, and 3.1$\sigma$ in the 1-3~MeV band.
Again, at higher energies only hints are obtained (see Table~\ref{tab:flux}). 

The most likely source location, the peak of maximum of likelihood
ratio distribution in the 1-3 MeV band of the combination of 7 VPs 
in 1995, is at l/b=311.5$^\circ$, -2.5$^\circ$. The error location 
contours are drawn from this distribution (see Fig.~\ref{fig:counterpart}).  

\subsection{Variability}
To check for possible source variability, we fit the fluxes  
of the individual VPs of the first five CGRO Phases
in the 4 standard bands with a constant flux.  
In the 1-3~MeV band is a $\chi^{2}$ value of 94.6 derived, 
which corresponds to a significance of 4.1$\sigma$ that the source
is variable, and -- subsequently -- to a probability of
$\sim$4.8$\times$10$^{-5}$ that the source is constant.
The flux variation is not significant in the other three energy
bands ($<$ 3$\sigma$).
Applying the same procedure to the light curves 
of the first five CGRO Phases (Fig.~\ref{fig:light-p15}),
i.e. averaging over relevant VPs, 
we derive a similar result: significant (3.9$\sigma$) flux variability
in the 1-3~MeV band, and insignificant one in the others.  

As was pointed out the $\chi^{2}$ statistics is statistical only
and therefore does not account for any systematic errors which might be 
inherent in the data (e.g. instrumental trends).
To overcome this uncertainty it was suggested (e.g. Zhang et al., 2000;
Torres et al., 2001a) to normalize the measured variability to the 
measured variability of sources, which are considered to 
be non-variable.
To do so, we calculated the so-called {\it I} variability index
(for definition and details see Zhang et al. (2000),
and Torres et al. (2001a)) 

\begin{equation}
 I = {\frac{\mu_{\rm{source}}}{\mu_{\rm{ref.\ sources}}}}
\end{equation}

which establishes how variable a source is with 
respect to a given source population.
Using the data of the first 5 CGRO Phases (Fig.~3) and compare them
to the same type of data for the prominent and rather stable MeV sources 
Crab (pulsar + nebula), Cyg~X-1, and 3C~273 
(e.g. Collmar et al., 1999), we derive an I index
of 5.2 in the 1-3 MeV band, which is just above the 3$\sigma$-level
for variability (Torres et al., 2001b).
Because Crab is the only significantly detected pulsar in 
this band, we added the X-ray binary Cyg X-1 and the quasar
3C~273 to the reference sample. These two source types are a
priori not considered to be constant in \gray\ flux, which 
-- in fact -- even strengthens the I index given above.  

The highest source fluxes are obtained in CGRO Phase 4.
Fig.~\ref{fig:light-p4} shows the 1-3~MeV light curve covering this period.
There are 2 VPs, 402 and 402.5 (October 18 - November 1, 1994), 
prior to the combination of the 7 VPs where the source was
most prominent.  The source is not detected in these 2 VPs,  
which proves variability on a time scale of less than 1 year, 
because the source rose from a non- to 7.2$\sigma$-detection
in the 1-3~MeV band.  The closest follow-up observation, VP 516.1 
(March 18-21, 1996), by COMPTEL was carried out 8 months later. 
It provides a non-detection of the source in the
1-3 MeV band (Fig.~\ref{fig:light-p4}).  
The 1-3~MeV long-term (CGRO Phase 1 to 5) light curve shows that
the source flux peaks overall in Phase 4 (Fig.~\ref{fig:light-p15}).

\subsection{Energy Spectra}
The energy spectrum of the source in VP 424 is consistent with 
a simple power-law shape of photon index 2.61$^{+0.33}_{-0.30}$. 
For the combination of the 7 VPs (414 - 424) we also 
derive a soft spectrum of photon index 2.53$^{+0.23}_{-0.20}$ 
(Fig.~\ref{fig:spectra}). The reduced $\chi^{2}$-values are 1.39
 and 0.52, respectively. 
The 1$\sigma$ errors are obtained by adding 1.0 to the minimum $\chi^{2}$-value.
The spectra indicate that the source
is weak at higher \gray\ energies (e.g. EGRET band),
and -- if no spectral turnover occurs -- is bright at hard X-rays.

\section{Discussion}

Previously, no \gray\ source was known at the position of this MeV source.  
The closest known \gray-emitting object, the EGRET source 3EG J1410-6147, is 
located outside the 4$\sigma$ error region. This source coincides 
spatially with the COS~B source 2CG~312-1 (Hermsen et al., 1977), 
and is tentatively identified with the supernova remnant G312.4-0.4 (Sturner \&
Dermer, 1995).
To find possible counterparts, we searched the SIMBAD and NED databases for
potential emitters of $\gamma$-rays near the source location.
The obtained possible counterparts are overlaid on the COMPTEL error contours
in Fig.~\ref{fig:counterpart}.

There are 46 X-ray sources located within the 4$\sigma$ error location contour.
With 2 exceptions, 2S~1417-624 and GS~1354-645, all of them are not identified
at other wavelengths. 
2S~1417-624 is a transient Be/X-ray binary pulsar system,
with a Be star of 14 and a neutron star of 1.4 solar masses.
BATSE observed a huge outburst of this source at hard X-rays
(pulsations up to 100 keV were detected), which started on 
August 26, 1994 and lasted $\sim$110 days  (Finger et al., 1996).
COMPTEL observed the source twice (VPs 402, 402.5) during this 
period but did not detect it. During the next $\sim$200 days, 
2S~1417-624 showed 5 smaller outbursts (Finger et al., 1996), which coincide in time 
with the significant COMPTEL detection of the new MeV source in 1995.
During the huge outburst in hard X-rays, the pulsed luminosity was found to be much less
than it would be estimated from the spin-up rate. This implys that most of
the power output is either unpulsed or outside of the hard X-ray range 
(Finger et al., 1996). Assuming 2S~1417-624 as counterpart of the MeV source, 
its 0.75-10 MeV luminosity would be 2.5$\times$10$^{37}$ erg s$^{-1}$ by using 
its upper-limit distance of 11.1 kpc. 
This luminosity is comparable to the pulsed one of
$\sim$2.2$\times$10$^{37}$ erg s$^{-1}$. 2S~1417-624 as counterpart would give an 
anticorrelation of the X- and $\gamma$-ray emission.

Cen X-3 is the only neutron star
XRB pulsar system for which \gray\ emission above 1~MeV was detected by CGRO. 
The significant detection of Cen X-3 by 
EGRET in October 1994 (Vestrand et al., 1997) proves that such objects 
could be occasional \gray\ emitters.
In fact, Romero et al. (2001) have proposed the
association of the $\gamma$-ray source 3EG J0542+2610 with the Be/X-ray transient
A0535+26. 
They assume that the \grays\ are produced by the impact of
relativistic protons (accelerated in the pulsar's magnetosphere)
into the accretion disk. 
This, subsequently, generates \grays\ around 67.7~MeV from 
$\pi_{0}$ decay, and also lower-energy \grays\ due to bremsstrahlung and 
Inverse-Compton scattering of relativistic secondary electron-positron 
pairs, which are produced by the decay of the charged pions.
Because the source also generates a strong X-ray field, the high-energy 
\grays\ (EGRET band) could be absorbed, whereas the MeV \grays\ could 
escape.   

GS~1354-645 is a transient black-hole X-ray binary (XRB) system. 
MeV emission was detected by COMPTEL from 2 such systems: Cyg~X-1, which shows
a persistent MeV flux (e.g. McConnell et al., 2000), and GRO~J0422+32, which
showed a transient MeV flux and was detected 
only once at MeV energies, during a strong outburst in hard
X-rays (van Dijk et al., 1995). 
However, no X-ray observations of GS~1354-624 during the activity period 
of the new COMPTEL source are reported.
The MeV spectrum of the new source is soft (Fig.~\ref{fig:spectra}) and 
therefore is consistent in shape with the COMPTEL spectra of Cyg~X-1
and GRO~J0422+32.

A weak galactic microquasar or microblazar might also be the counterpart of
this new MeV source. Recently Kaufman Bernad$\acute{o}$ et al. (2002)
published a model for variable
$\gamma$-ray emission from such objects. Relativistic electrons of
a precessing jet upscatter stellar photons from the companion 
to \grays\, thereby generating a variable \gray\ emission.
Although the model is tuned to GeV \grays, a MeV source can be obtained 
by assuming a different spectrum for the relativistic electrons. 
Because the inverse-Compton emission is beamed but the disk emission
is not, any of the unidentified (galactic) X-ray sources could
 -- according to this scenario -- be the counterpart. 
Due to different beaming factors in radio (synchrotron) and \grays,
a strong radio source is not necessarily expected.      

Massive stars, supernova remnants, high velocity clouds (HVC),
and pulsars are also accelerating particles to relativistic 
energies by shocks or strong electromagnetic fields.
Therefore such objects are also potential $\gamma$-ray emitters.
Two HVCs are located within the 4$\sigma$ error contours. 
Gamma-ray emission from HVC was reported by Blom et al. (1997). 
The time scale for flux variability of less than 1 year would provide
constraints on the emission site in HVC.
SIMBAD gives 3 massive stars, however no SNR  
within the 4$\sigma$ error contour. 
Two of them are Wolf-Rayet stars 
(WC9 for LSWR 9 (Smith, 1968), WN6 for WR 61 (Torres-Dodgen $\&$ Massey, 1988)),
and the third one (HD~123008) is a O9.5I (Cruz-Gonzalez et al., 1974). 
Only HD 123008 has a photometric distance of 5.83 kpc 
(Cruz-Gonzalez et al., 1974). There is no detection of non-thermal
radio emission for these three massive stars.
Two radio pulsars are located within
the 3$\sigma$ error contour. They are not likely counterparts of the MeV
source because the detected $\gamma$-ray pulsars do not show obvious time
variability and typically have harder spectra (index $>$ -2.15).

We investigated also extragalactic objects.
There is only one AGN, the Circinus Galaxy, located within the 4$\sigma$ 
error contour. It is a Seyfert type 2 object at a distance of $\sim$3~Mpc.
Seyfert galaxies were observed by CGRO, 
in particular by the OSSE experiment, to have a strong spectral cut off
around 100~keV (e.g. Johnson et al., 1997). Subsequently no Seyfert has yet
been detected at 1~MeV (e.g. Maisack et al., 1995) or above, which makes
the Circinus AGN unlikely to be the counterpart. 
Torres et al. (2002) recently showed that by gravitational microlensing
a distant \gray\ emitting blazar could be a strong and variable 
\gray\ source in the observer's frame without a radio or optical 
counterpart. According to their calculations is the magnification 
factor energy dependent, in the sense that it becomes stronger
towards lower energies (their calculations covered the band 
between 100 MeV and 10 GeV). This could mean, that a variable 
MeV source appears in the observer's frame without a visible 
GeV source. This scenario might provide a possibility to explain
our transient MeV source by an extragalactic object although 
no obvious extragalactic object is within the error box.

28 X-ray sources are embedded in the Circinus galaxy. Some of them 
are likely to be XRBs (Sambruna et al., 2001; Bauer et al., 2001).
Placing the MeV source at this distance, a 0.75-10 MeV luminosity of 
$\sim$1.8$\times$10$^{42}$ erg s$^{-1}$ is implied for the activity period
in VPs 414-424. 
Assuming Eddington accretion, the black-hole mass in an XRB would have to 
be in excess of $\sim$ 10$^{4}$ M$_{\odot}$. Therefore, considering extragalactic 
XRBs in the nearby galaxy as the counterpart will result in a 
super-luminosity problem, unless the source would be a microquasar or microblazar,
for which jet emission is assumed (e.g. Markoff et al., 2001). 
Relativistic beaming could significantly enhance the apparent luminosity.

The MeV source resembles a 
special subgroup of unidentified EGRET sources located at low galactic
latitudes ($|$b$|$ $<$ 10$^{\circ}$).
These low-latitude sources show a highly variable flux, a soft \gray\ spectrum,
and do not spatially coincide with any known potential $\gamma$-ray emitters
(Torres et al., 2001a).
Torres et al. (2001a) suggest that isolated Kerr-Newman black holes
could be the objects behind. One of them, 3EG~J1828+0142, was modeled
by Punsly et al. (2000) based on that theory. 
Apart from time variability and soft spectrum, the model predicts
in addition a broad spectral peak at MeV energies.
Unfortunately, no further measurements, in particular at neighboring
energies are available, to check whether the observed MeV emission is
a peak or hump-like feature. 

In summary,  searching the NED and SIMBAD databases provides no obvious 
counterpart within the 4$\sigma$ error contour. The source properties, flux 
variability and steep spectrum, resemble a special group of
unidentified EGRET sources, which are also located near the galactic plane. 
However, XRB pulsars or microquasars or blazars could 
also generate such an emission profile.
 
\section{Conclusion}

A new COMPTEL $\gamma$-ray source is significantly detected at MeV energies
between 0.75 and 3~MeV. The source is variable on a time scale of
less than one year. The MeV spectrum for the high-state period
is soft and can be represented by a simple power-law shape. Due to 
the absence of any simultaneous observations at other wavelengths,   
no identification is possible. Future contemporaneous X- and
$\gamma$-ray observations have to be awaited
to probe the nature of this source.

\begin{acknowledgements}
This research was supported by the German government through DLR grant 50 QV 9096 8, by NASA under contract NAS5-26645, and by the Netherlands
Organization for Scientific Research NWO. 
This research has made use of the SIMBAD database operated at CDS, Strasbourg, France, and the NASA/IPAC Extragalactic Database (NED) which is 
operated by the Jet Propulsion Laboratory, California Institute of Technology, under contract with the National Aeronautics and Space Administration.
The authors would like to thank the anonymous referee for the helpful comments.
\end{acknowledgements}

\end{document}